\documentclass[iop,useAMS]{emulateapj}

\usepackage{graphicx}
\usepackage[percent]{overpic}
\usepackage{multirow}
\usepackage{color}
\usepackage{rotating}

\def\ltsima{$\; \buildrel < \over \sim \;$}
\def\simlt{\lower.5ex\hbox{\ltsima}}
\def\gtsima{$\; \buildrel > \over \sim \;$}
\def\simgt{\lower.5ex\hbox{\gtsima}}
\def\kms{{\rm\,km\,s^{-1}}}
\def\kpc{{\rm\,kpc}}
\def\mpc{{\rm\,Mpc}}


\def\deg{^\circ}

\def\ltsima{$\; \buildrel < \over \sim \;$}
\def\gtsima{$\; \buildrel > \over \sim \;$}

\shorttitle{Satellite-satellite correlations}
\shortauthors{Ibata et al.}

\begin{document}

\title{Eppur si muove:\\ Positional and kinematic correlations of satellite pairs in the low Z universe}

\author{Rodrigo A. Ibata\altaffilmark{1}}
\author{Benoit Famaey\altaffilmark{1}}
\author{Geraint F. Lewis\altaffilmark{2}}
\author{Neil G. Ibata\altaffilmark{3}}
\author{Nicolas Martin\altaffilmark{1,4}}

\altaffiltext{1}{Observatoire astronomique de Strasbourg, Universit\'e de Strasbourg, CNRS, UMR 7550, 11 rue de l'Universit\'e, F-67000 Strasbourg, France; rodrigo.ibata@astro.unistra.fr}
\altaffiltext{2}{Sydney Institute of Astronomy, School of Physics A28, University of Sydney, NSW 2006, Australia}
\altaffiltext{3}{Trinity College, Trinity Street, Cambridge, CB2 1TQ, United Kingdom}
\altaffiltext{4}{Max-Planck-Institut f\"{u}r Astronomie, K\"{o}nigstuhl 17, 69117 Heidelberg, Germany}

\begin{abstract}
We have recently shown (Ibata et al. 2014) that pairs of satellite galaxies
located diametrically opposite each other around their host possess
predominantly anti-correlated velocities. This is consistent with a scenario
in which $\simgt 50$\% of satellite galaxies belong to
kinematically-coherent rotating planar structures. Here we extend this
analysis, examining satellites of giant galaxies drawn from an SDSS
photometric redshift catalog.  We find that there is a $\sim 17$\%
overabundance ($> 3 \sigma$ significance) of candidate satellites at
positions diametrically opposite a spectroscopically confirmed satellite.
We show that $\Lambda$CDM cosmological simulations do not possess this
property when the contamination is included.  After subtracting
contamination, we find $\sim 2$ times more satellites diametrically opposed
to a spectroscopically confirmed satellite than at $90\deg$ from it, at
projected distances ranging from 100 to $150\kpc$ from the host.  This
independent analysis thus strongly supports our previous results on
anti-correlated velocities.  We also find that those satellite pairs with
anti-correlated velocities have a strong preference ($\sim 3:1$) to align
with the major axis of the host whereas those with correlated velocities
display the opposite behavior. We finally show that repeating a similar
analysis to Ibata et al. (2014) with same-side satellites is generally hard
to interpret, but is not inconsistent with our previous results when strong
quality-cuts are applied on the sample. This addresses all concerns recently
raised by Cautun et al. who did not uncover any flaw in our previous
analysis, but may simply have hinted at the physical extent of planar
satellite structures by pointing out that the anti-correlation signal
weakens at radii $>150\kpc$. All these unexpected positional and
kinematic correlations strongly suggest that a substantial fraction of
satellite galaxies are causally-linked in their formation and evolution.\end{abstract}


\section{Introduction}
\label{sec:Introduction}

Within the framework of the standard $\Lambda$CDM model of cosmology, a consensus has built up from the observation of the kinematics and stellar populations of satellite galaxies \citep{1998ARA&A..36..435M,2007ApJ...670..313S,2007MNRAS.380..281M,Diemand:2007hb}, and from detailed modeling of the formation of cold dark matter (CDM) structures \citep{Moore:1999ja,Bullock:2000bn,2004MNRAS.355..819G}, that satellite galaxies are embedded in massive dark matter sub-halos, although the vast majority of such sub-halos (and possibly many of higher mass than those that harbor faint satellites) remain entirely devoid of stars and gas \citep{2010AdAst2010E...8K}.

It is in this context that the detection of coherent planes of satellites in the Milky Way \citep{1976MNRAS.174..695L,1976RGOB..182..241K,2005A&A...431..517K,2008ApJ...680..287M,2012MNRAS.423.1109P,2013MNRAS.435.2116P} and Andromeda \citep{2007MNRAS.374.1125M,2013Natur.493...62I,2013ApJ...766..120C} (M31) galaxies is surprising and especially interesting\footnote{The non-isotropic distribution of known satellite galaxies of the Milky Wat was first noted at a time when the current cosmological model did not exist yet \citep{1976MNRAS.174..695L,1976RGOB..182..241K}, but its likely conflict with $\Lambda$CDM was raised for the first time by \citet{2005A&A...431..517K}}. Both of these Local Group systems possess very thin planes of satellites with coherent kinematic properties. The Milky Way has a striking structure in that it appears that most satellites and clusters beyond about 10 kpc are in a polar planar structure, which appears to be rotating about the Milky Way\footnote{A similar phase-space correlated system is also found around M81 \citep{2013AJ....146..126C}.} \citep{2013MNRAS.435.2116P,2014MNRAS.442.2362P,Kim:2014ui}. The case of M31 is perhaps even more striking as our vantage point from outside the M31 system actually gives us a clear panoramic view that allows an easy interpretation of the observations: approximately $50$\% of the satellites surrounding that galaxy belong to a planar structure that is very thin ($12.6\kpc$ rms) but very extended ($\sim 400\kpc$ diameter), and that possesses coherent kinematics suggestive of common rotation about M31 \citep{2013Natur.493...62I}. Selecting systems from the Millennium II simulation (MS2) of galaxy formation and evolution \citep{2009MNRAS.398.1150B,2013MNRAS.428.1351G} that are similar in mass and environment to M31 shows that such alignments of satellites are extremely rare, occurring around only $0.03$--$0.04$\% of the simulated galaxies (\citealt{2014ApJ...784L...6I}, see also \citealt{2014MNRAS.442.2362P}). The Milky Way structure is equally unlikely to be found in such simulations \citep{2010A&A...523A..32K}.

The presence of satellite alignments in both the Milky Way and M31, which a-priori one would expect to be statistically independent systems, therefore appears to place a tension on $\Lambda$CDM models, suggesting that there is a missing ingredient in such galaxy formation simulations.  

One particular property of the M31 plane of satellites is that it points almost directly towards us (to within $1\deg$). Given this very peculiar orientation, it is natural to wonder whether the Local Group giant galaxies are at all representative of the larger population\footnote{In fact, some alternative scenarios could perhaps provide a natural explanation for this special orientation of the M31 planar structure \citep{2013A&A...557L...3Z}.}. With these doubts in mind, we set out in \citet[][hereafter Paper~I]{2014Natur.511..563I} to ascertain whether similar galaxy alignments could be found in a sample of more distant galaxies. We developed a simple test that could be applied to systems with well-measured velocities consisting of a host galaxy surrounded by at least two satellites. If such systems possess the coherent rotation-like motion observed in M31 and the Milky Way, satellites on opposite sides of the host will in general display anti-correlated velocities. This signal can be made stronger by selecting samples that are more likely to be edge-on, which can be achieved by choosing satellites that are diametrically opposite each other and that possess significant velocity differences with respect to their host. 

We found that with the parameters that gave the sample of highest significance ($4\sigma$), the ratio of anti-correlated to correlated satellite pairs is $10:1$. This is consistent with a scenario in which $\simgt 50$\% of satellites follow planar configurations. Furthermore, the direction defined by the satellite pairs was found to correlate with the surrounding large scale structure out to $\sim 2\mpc$ ($\sim 7\sigma$ significance). Despite these strong detections, a $4\sigma$ significance means that we might just have been unlucky (with a bit more than one in a $\sim 15$ thousand chance) due to the limited number of suitable systems in the spectroscopic samples: with the highest contrast selection we found only  21 diametrically opposed satellite pairs with anti-correlated velocities out of a total of 23 pairs within a tolerance angle of $8\deg$ from each other (see Paper I). Similar concerns have recently been raised by \citep[][hereafter C14]{Cautun:2014un}.

The aim of the present contribution is first to address the concern of the low statistics in the analysis presented in Paper~I. Here we will repeat that analysis, but instead of using pairs of spectroscopically-confirmed satellites, we will require only a single spectroscopic satellite (thus enhancing the sample of suitable hosts by an order of magnitude, and extending the redshift upper limit of the hosts to $z=0.1$). We will then examine the incidence of objects in the SDSS photometric catalog as a function of angle from the spectroscopically-confirmed satellite. This is similar to what was done in C14, except that we make explicit use of photometric redshift information to reduce the background contamination. The robustness of our contamination estimation will then be checked through the equivalent angular distribution of contamination-free spectroscopically-confirmed pairs of satellites out to $z=0.05$.

Secondly, we will study the distribution of the satellite pairs in relation to the projected major axis of the host galaxy. Several studies have shown that there is a clear excess of satellites at angles near the host major axis \citep{2004MNRAS.348.1236S,2009MNRAS.395.1184S,2005ApJ...628L.101B,Yang:2006jp,2007MNRAS.376L..43A,2010ApJ...709.1321A}. The effect is detected in both elliptical and spiral galaxy samples, and is strongest for red, high-mass satellites. These correlations are consistent with models in which host ellipticals are oriented in the same direction as their dark matter halos, while host spirals have the same angular momentum axis as their halos \citep{2010ApJ...709.1321A}. We are particularly interested in establishing the link, if any, between these satellite-host correlations and the satellite-satellite correlations discovered in Paper~I.

Finally, we will reply to other concerns raised in a recent paper (C14) that claims that the results presented in Paper I are not robust. 

The layout of this article is as follows: the selection of the galaxy sample with photometric redshifts is presented in Section~\ref{sec:Sample_Selection}; Section~\ref{sec:Satellite_satellite} presents the analysis of the satellite-satellite correlations; the angular distribution in relation to the major axis is presented in Section~\ref{sec:major_axis}; and we discuss and draw conclusions from our work in Sections~\ref{sec:Discussion} and \ref{sec:Conclusions}, respectively.

\begin{figure}
\begin{center}
\includegraphics[viewport= 30 30 530 535, clip,width=\hsize]{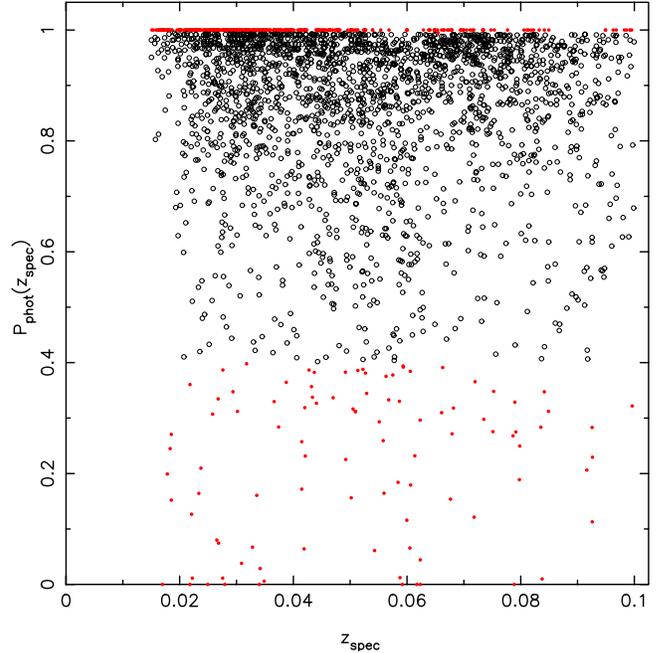}
\end{center}
\caption{Probability of the photometric redshift, for the spectroscopically-confirmed ``primary'' satellites. The probability corresponds to the sum over the three nearest redshift bins around the (spectroscopically-measured) redshift of the host. Open circles mark the objects selected as reliable satellite candidates. For comparison, the fraction of all objects in the \citet{2012ApJS..201...32S} photometric redshift catalog that have $0.4 < P_{\rm phot}(z_{\rm spec}<0.126) < 0.995$ is 8\%, implying that $\sim 92$\% of the background population will be removed with this $P_{\rm phot}(z_{\rm spec})$ selection.}
\label{fig:redshift_comparison}
\end{figure}

\eject

\section{Sample Selection}
\label{sec:Sample_Selection}

As in Paper~I, we select host galaxies from the SDSS ``NYU Value-Added Galaxy Catalog'' \citep{2005AJ....129.2562B} (their VAGC-DR7 catalogs) that have magnitudes in the range $-23\le {\rm M_r} \le -20$. (Throughout this work, physical quantities are converted to be consistent with the cosmological parameters derived by the Planck mission, \citealt{2014A&A...571A..16P}). We keep only those objects that are relatively isolated, having no brighter neighbor within a projected distance of $0.5\mpc$ and $1500\kms$. These criteria are intended to produce galaxies roughly comparable in magnitude and environment to M31 and the Milky Way. For comparison, M31, the Milky and the LMC have V-band absolute magnitudes of ${\rm M_V=-21.1}$, ${\rm M_V=-20.6}$, and ${\rm M_V=-18.1}$, respectively \citep{1998gaas.book.....B}.

Around each one of these hosts, we next select high confidence satellites from the SDSS spectroscopic sample, as follows. We will refer to these as ``primary'' satellites. These objects are required to have magnitudes ${\rm M_r}<-16$, to be at least one magnitude fainter than the host, and to lie at a projected radius between $20\kpc$ and $150\kpc$ (these were the selection criteria defined in Paper~I). We also apply the same velocity difference criterion as in Paper~I, namely that the ``primary'' satellite must have a minimum velocity difference greater than  $|v-v_{\rm host}|_{\rm min} = \sqrt{2} \times 25\kms$, and a maximum velocity difference of $|v-v_{\rm host}|_{\rm max} = 300 \exp[-(R/300\kpc)^{0.8}]\kms$. The former criterion ensures that any planar system is more likely to be detected edge-on rather than face on, while the latter is an envelope relation chosen to reduce interlopers (both choices are explained in detail in Paper~I). We limit the sample to $z>0.002$, to avoid some clearly dubious measurements and contamination, and extend the upper redshift range to $z=0.1$.

\begin{figure}
\begin{center}
\includegraphics[viewport= 30 30 530 535, clip,width=\hsize]{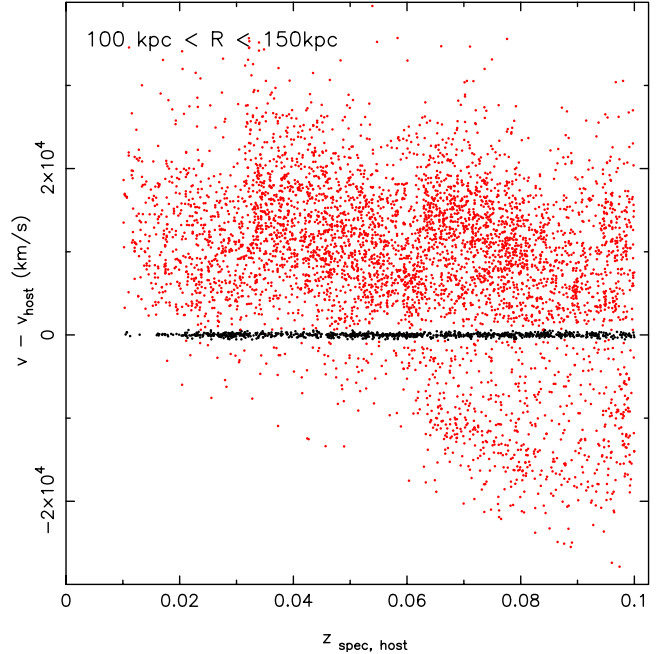}
\end{center}
\caption{Contamination in the photometric redshift sample. Here we display the galaxies with photometric redshifts consistent with being satellites of a host (i.e. objects having $0.4 < P_{\rm phot}(z_{\rm spec, host}) < 0.995$) within the radial range $100<R<150\kpc$, and that also have a measured spectroscopic redshift. Only 15.5\% of these galaxies turn out to have velocities within $600\kms$ of their host (a typical escape velocity from a Milky Way type system), and can be taken to be bona-fide satellites of the considered hosts. These satellites are marked in black. Contaminants (red dots) comprise 84.5\% of the sample.}
\label{fig:contamination}
\end{figure}

As discussed above, Paper~I demonstrated the velocity anti-correlation of satellite pairs on opposite sides of their hosts. The requirement that the satellites should not be on the same side of the host galaxy was a conservative choice implemented to avoid the inclusion of binary satellite systems (which are necessarily statistically correlated) in our sample, and also to avoid the massive satellites influencing each other's kinematics (this is discussed in more detail in \S\ref{sec:cautun} below). For the same reason, here we first only consider pairs separated by more than $90\deg$ around the host. To match to our bright ``primary'' satellites, we select ``secondary'' candidate satellites from the SDSS photometric redshift catalog by \citet{2012ApJS..201...32S}. This catalog provides a photometric redshift probability distribution (tabulated at 35 bins between $z=0$ and $z=1.1$) for galaxies observed in the main SDSS photometric survey. In Figure~\ref{fig:redshift_comparison} we display for the primary satellites $P_{\rm phot}(z_{\rm spec})$, the sum of the probability values of the correct (spectroscopic) redshift bin plus the two adjacent bins (each redshift bin has width $\delta z = 0.031$). Comparison to the spectroscopic redshifts shows that using the photometric redshift probability from only the highest value bin gives less reliable estimates. We reject those galaxies that have $P_{\rm phot}(z_{\rm spec})<0.4$; while this only removes 1\% of bona-fide satellites, it excludes $\sim 92$\% of background contaminants. We also decided to reject those objects with $P_{\rm phot}(z_{\rm spec})>0.995$, as there is a clear discontinuity in the distribution of $P_{\rm phot}(z)$ near unity, and this suggests to us that these photometric redshift solutions (14\% of the sample) are likely unreliable.

Whereas a total of only $380$ satellite galaxy pairs passed the selection criteria set out in Paper~I, we find 6355 ``primary'' satellites around 5661 different bright hosts with these selections.

Of course, some fraction of the sources selected on photometric redshifts will not belong to the host system under consideration despite the imposed $P_{\rm phot}(z_{\rm spec})$ criterion. We checked this contamination fraction by using those sources that also possess SDSS spectroscopic redshifts. In Figure~\ref{fig:contamination} we show the candidate satellites chosen in a distance range of $100<R<150\kpc$, and that have photometric redshifts that are consistent with belonging to the host system (i.e. $0.4 < P_{\rm phot}(z_{\rm spec, host}) < 0.995$). Evidently, the majority of these sources are contaminants (red dots), with only 15.5\% having velocities within $600\kms$ of the host galaxy (which we consider to be a reasonable estimate of the escape velocity of such systems). Any statistical test based on satellites selected by photometric redshift will need to take this large contamination into account.

\begin{figure}
\begin{center}
\includegraphics[viewport= 30 30 560 535, clip,width=\hsize]{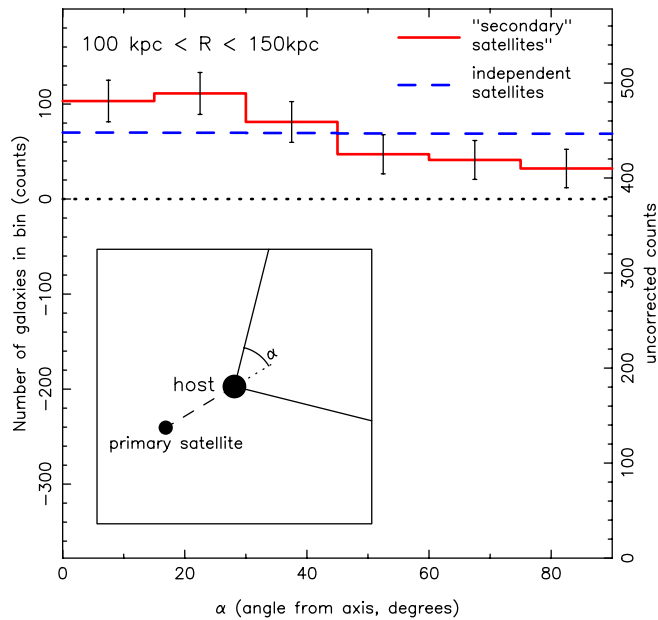}
\end{center}
\caption{Counts of ``secondary'' satellites as a function of angle. The insert shows the definition of the angle $\alpha$. Having selected a bright host galaxy and a ``primary'' satellite from the SDSS spectroscopic catalog, we count the number of candidate (``secondary'') satellites listed in the \citet{2012ApJS..201...32S} photometric redshift catalog that lie at an angle $\alpha$, within a projected distance of $100\kpc < R < 150\kpc$. There is a clear overdensity of galaxies at $\alpha<45\deg$. In contrast, if the satellite positions were determined by their relation to the host major axis (Fig.~\ref{fig:primaries_host_axis}), and were otherwise statistically independent of each other, we would expect the almost flat distribution shown with the dashed line. A Kolmogorov-Smirnov test shows that these two distributions are different (99.8\% confidence).}
\label{fig:angle}
\end{figure}

\begin{figure}
\begin{center}
\includegraphics[viewport= 30 30 530 535, clip,width=\hsize]{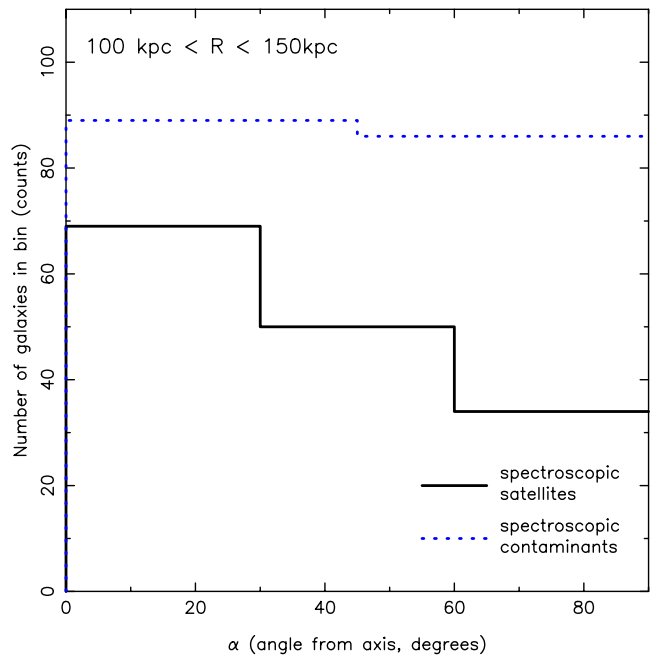}
\end{center}
\caption{As Figure~\ref{fig:angle}, but using only spectroscopically selected satellites. The ``secondary'' satellites here have velocities within $600\kms$ of their hosts (black histogram), i.e. the selection is designed to be comparable to the corrected counts in Figure~\ref{fig:angle}. The strong tendency for the satellites to be located diametrically opposite each other that was revealed in Figure \ref{fig:angle} is confirmed by this cleaner sample, in which the ratio of the inner to outer bins is $2.0\pm0.4$. The sample is limited to $z=0.05$, to avoid requiring the putative structures to contain two very bright satellites (as we detail in Section~\ref{sec:Parameter_selections} below, none of the satellite pairs identified in Paper~I could have been detected beyond $z=0.06$). The blue dotted histogram shows the angular distribution of spectroscopic contaminants for $z<0.1$ (these are galaxies having $|v-v_{\rm host}|>600\kms$).}
\label{fig:angle_spectro}
\end{figure}

\section{Satellite-satellite correlations}
\label{sec:Satellite_satellite}

\subsection{Background contaminated sample}
If the analysis presented in Paper~I is correct, the velocity criteria applied in the present work to select primary satellites should have enhanced the probability of finding any planar rotating alignments in a more edge-on orientation. With such a geometric configuration, the conclusions of Paper~I would predict a higher incidence of ``secondary'' satellites at locations opposite the primary. This result is confirmed by the angular distribution of satellites displayed in Figure~\ref{fig:angle}, which, for the radial range $100<R<150\kpc$, shows a peak of higher counts out to an angle $\alpha$ of $45\deg$, followed by a flat plateau. (We chose $100<R<150\kpc$ for this initial exploration, as it corresponds to a radial region where signal was previously found in Paper~I, and as we show below, it is sufficiently distant from the hosts that the contamination does not vary strongly over the range. It also corresponds to the region in Figure~\ref{fig:ratio} below where the signal appears strongest). Evidently, there is a substantial enhancement of galaxies at $\alpha<45\deg$, amounting to $3.4\sigma$ over the counts at $\alpha>45\deg$.

\subsection{Accounting for contamination}
\label{subsec:Contamination}

To uncover the intrinsic angular distribution of satellite galaxies, we need to correct the counts shown in Figure~\ref{fig:angle} for the contaminating populations. By comparing the photometrically-selected satellite candidates to spectroscopically-confirmed galaxies in the same radial range from their hosts, we found previously in  Figure~\ref{fig:contamination} that the contamination fraction is 84.5\%. 

In order to test this contamination estimate, we implemented an alternative method, shifting the host galaxies and their primary satellites by two virial radii along the North, South, East and West cardinal directions. (Following C14, we assumed that hosts with $M_r<-22.5$ have virial radii of $r_{\rm vir} = 500\kpc$, those with $-22.5 \le M_r \le -21.5$ have $r_{\rm vir} = 315\kpc$, and those with $M_r>-21.5$ have $r_{\rm vir} = 150\kpc$). We find that in the same radial range ($100<R<150\kpc$), the number of secondary satellites within $0<\alpha<90\deg$ in the offset locations amount to 78\% of the counts at the true locations. However, this contamination fraction is likely underestimated since galaxies are clustered, and by probing away from the true host locations, we will on average probe lower density regions (especially in projection). Thus we interpret this as a confirmation that the contamination estimate from the comparison to bona-fide spectroscopic satellites is reliable.

We therefore corrected the galaxy counts in Figure~\ref{fig:angle} using the 84.5\% contamination derived in Figure~\ref{fig:contamination} for the same radial range (the uncorrected counts are displayed on the right-hand axis). This suggests that there are approximately three times more ``secondary'' satellite galaxies opposite a ``primary'' satellite than at $90\deg$ to the ``primary''. We were surprised by this very strong correlation, and suspecting an error, we immediately checked to see if the tendency could be corroborated with spectroscopically-measured galaxies. The result is shown in Figure~\ref{fig:angle_spectro}, which constitutes an identical experiment to that displayed in Figure~\ref{fig:angle}, except that the ``secondary'' satellites are genuine members of the system as they are required to have spectroscopic velocities within $600\kms$ of the host (and they extend out to only $z=0.05$). The factor of $2.0\pm0.4$ higher counts in the lowest $\alpha$ bin compared to the highest $\alpha$ bin confirms the result shown in Figure~\ref{fig:angle}. Satellite galaxies in the $100<R<150\kpc$ range appear to possess a strong tendency to lie diametrically opposite each other. In contrast, the contaminants (galaxies with velocities $|v-v_{\rm host}|>600\kms$) are consistent with a flat angular distribution (blue dotted histogram).

Obviously compared to the sample derived in Paper~I using spectroscopic velocities, the present selection based on photometric redshifts is much more contaminated by extraneous foreground and background galaxies, and as a result we cannot realistically expect to impose a small tolerance angle ($\alpha$) in the selection of the secondary satellites. This consideration naturally motivates simply using $\alpha=45\deg$ and comparing the galaxy counts in the quadrant marked ``O'' (for opposite) in the sketch presented in Figure~\ref{fig:ratio}) to the counts in the regions marked ``A'' (for adjacent). The ratio of these counts as a function of distance is also displayed in Figure~\ref{fig:ratio}. In the radial range $30\kpc$ to $150\kpc$, we find $N_A=2431$ and $N_O=2665$ (no correction is made for contamination), implying a $9.6$\% overdensity of galaxies in the opposite quadrant, statistically significant at the $3.3\sigma$ level. 

\subsection{Comparison with simulations}
It is interesting to compare this measured spatial correlation with expectations from the MS2 simulation. To this end we ``observed'' the MS2 simulation in a similar way to that done in Paper~I. As before, we selected isolated host galaxies (in the range $-23\le {\rm M_r} \le -20$) with no brighter neighbor within $0.5\mpc$. The simulation was placed at 1000 random distance locations out to $z=0.1$ (chosen fairly to take into account the increasing volume with $z$), and viewed from a random orientation. ``Primary'' satellites were selected with magnitudes brighter than $r=17.8$. As for the observations, these ``primary'' satellites must also have a minimum velocity difference of $|v-v_{\rm host}|_{\rm min} = \sqrt{2} \times 25\kms$, and a maximum velocity difference of $|v-v_{\rm host}|_{\rm max} = 300 \exp[-(R/300\kpc)^{0.8}]\kms$ with respect to their hosts. The ``secondary'' satellites are chosen to be brighter than $r=19.5$ (see below), and must have a maximum velocity difference of $|v-v_{\rm host}|_{\rm max} = 600\kms$. Both primary and secondary satellites must be at least one magnitude fainter than the host, but brighter than ${\rm M_r}=-16$. In contrast to Paper~I where we selected only the brightest two satellites around each host, here we accept all objects that satisfy these criteria. The \citet{2013MNRAS.428.1351G} MS2 catalog differentiates normal galaxies from so-called ``orphans'', these latter objects being systems whose parent subhalo is no longer resolved. We discard such ``orphan'' galaxies, as they are possibly tidally disrupted systems; in any case, C14 find that their rejection or inclusion does not significantly affect their angular correlation results.

The black histogram in Figure~\ref{fig:ratio_MS2} shows the $N_O/N_A$ ratio, as a function of distance, as measured directly from the simulations. This would be the prediction if there were no contaminating populations in the observations. However, as we demonstrated in Figure~\ref{fig:contamination}, the fidelity of the photometric redshifts for ensuring membership of a given host system is not high. We derived the contamination as a function of distance (blue-dotted line in Figure~\ref{fig:ratio_MS2}) from the overlap between the photometric redshift sample and the spectroscopic sample, using hosts of exactly the same properties as those used in the angular correlation measurements.

\begin{figure}
\begin{center}
\includegraphics[viewport= 30 30 530 535, clip,width=\hsize]{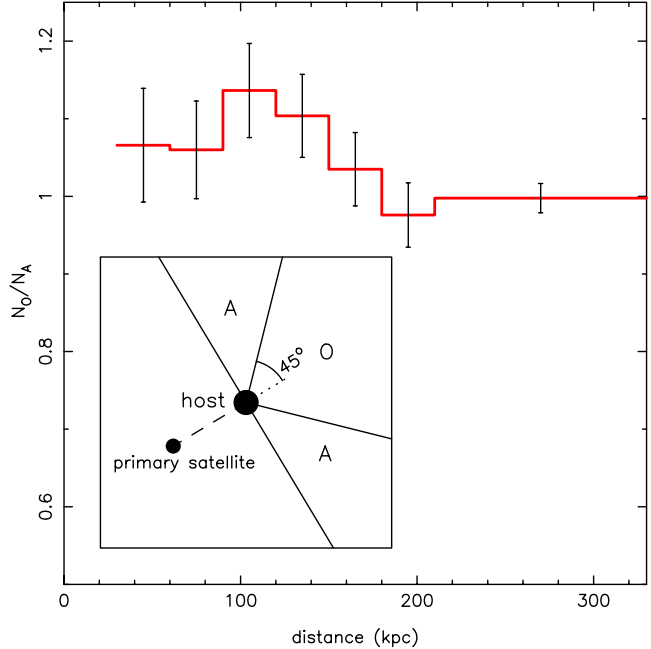}
\end{center}
\caption{Ratio of raw (uncorrected) galaxy counts of ``secondary'' satellites in the ``opposite'' to ``adjacent'' spatial regions. The sketch in the insert shows the implemented test. The statistic we use is the ratio of the number of candidate satellites within the opposite quadrant to the primary (marked ``O'') compared to the adjacent regions (marked ``A''). The ratio $N_O/N_A$ is high out to $\sim 180\kpc$. Summing between $30\kpc$ to $150\kpc$ we find a $3.3\sigma$ overabundance of galaxies in the ``O'' quadrant compared to the two ``A'' areas.}
\label{fig:ratio}
\end{figure}

The contaminants should be randomly distributed in $\alpha$, as the blue dotted histogram in Figure~\ref{fig:angle_spectro} shows that there is no angular dependence of the contamination fraction. Therefore the effect of the contaminants should be to add equal amounts of noise to the $N_O$ and $N_A$ counts. This means that the expected $N_O/N_A$ ratio should fall to a level close to unity, as shown in the red line histogram in Figure~\ref{fig:ratio_MS2}, which is the MS2 prediction for the observed distribution (red line histogram in Figure~\ref{fig:ratio}).
Thus both Figures \ref{fig:angle} and \ref{fig:angle_spectro} are revealing a spatial correlation property of satellite galaxies that is not predicted by the Millennium II simulation.

The contamination fraction we calculate, based on a comparison between the photometric and spectroscopic samples, is only strictly valid to $r=17.8$, yet the majority of the photometric redshift sample lies at fainter magnitudes. In practice however, our adopted redshift probability criterion $0.4 < P_{\rm phot}(z_{\rm spec, host}) < 0.995$ strongly curtails the faint sources, so that 86\% of the sample lies at $r<19.5$. We have checked that  qualitatively identical angular correlation results are obtained if we restrict the photometric redshift sample to $r<19.5$. Furthermore, while there may be some concern that we have thereby {\it underestimated} the contamination, the consequence would be that the expected MS2 profile of $N_O/N_A$ in Figure~\ref{fig:ratio_MS2} should be actually flatter and closer to unity than shown. Hence we do not consider our extrapolation of the contamination fraction to be a limitation in our analysis.

\begin{figure}
\begin{center}
\includegraphics[viewport= 30 30 530 535, clip,width=\hsize]{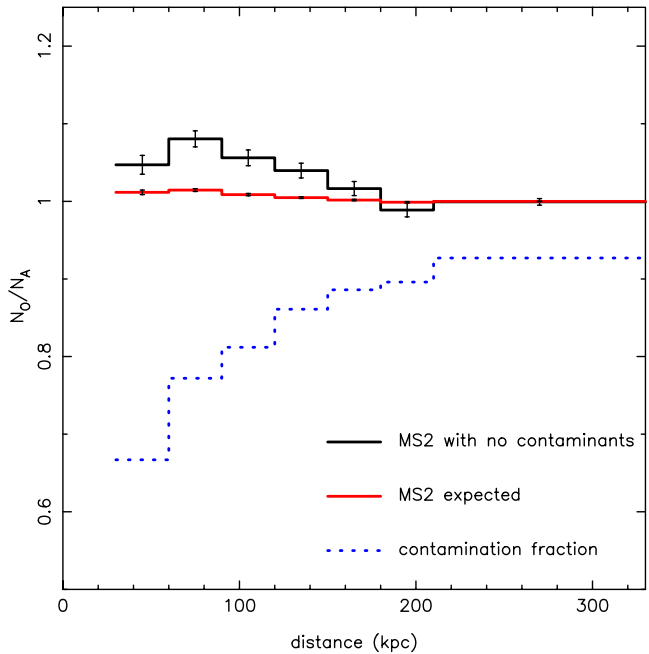}
\end{center}
\caption{The ratio of counts in the opposite and adjacent regions defined in Figure~\ref{fig:ratio} predicted by the Millennium II simulation. Satellites in the MS2 do show a tendency to lie opposite a ``primary'' satellite (selected according to the same magnitude, distance and kinematic properties applied to the SDSS ``primary'' satellites), so that $N_O/N_A>1$ over the entire distance range considered (black histogram). However, given the very large contamination fraction (blue dotted line) in our satellite sample derived from photometric redshifts, the expected ratio predicted for such observations is very close to unity (red histogram).}
\label{fig:ratio_MS2}
\end{figure}

\begin{figure}
\begin{center}
\includegraphics[viewport= 30 30 530 535, clip,width=\hsize]{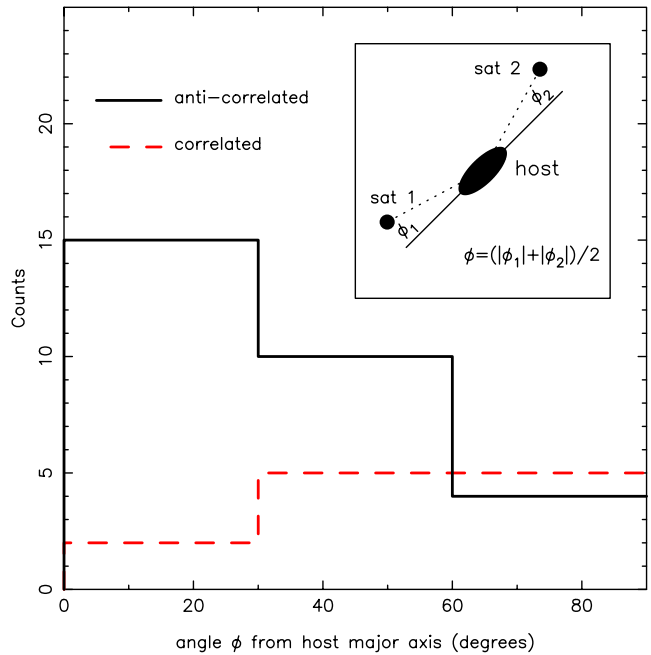}
\end{center}
\caption{$\phi$ distribution of satellite pairs with spectroscopic redshifts. The insert shows a sketch of the definition of the angle $\phi$ with respect to the host major axis. The sample is that of Paper~I, selected with tolerance angle $\alpha=15\deg$. The full line histogram shows the 29 satellite pairs with anti-correlated velocities that are consistent with belonging to thin planar rotating alignments of satellites. This sub-sample has $\langle \phi \rangle = 35.0\deg \pm 4.4\deg$. The sub-sample of pairs with correlated velocities shows the opposite tendency to be aligned closer to the projected minor axis of the host, with $\langle \phi \rangle = 57.2\deg \pm 6.1\deg$. (The directions of the host's projected major and minor axes are $\phi=0\deg$ and $\phi=90\deg$, respectively). Note that the planes found around the MW and M31 are, in this sense, exceptions.}
\label{fig:host_axis}
\end{figure}

\section{Correlation of satellite pairs with host major axis}
\label{sec:major_axis}

We next examine possible correlations with the direction defined by the projected major axis of the host galaxies. We start with the sample of 41 systems defined in Paper~I selected using a tolerance angle of $\alpha=15^\circ$. For each of these spectroscopically confirmed satellite pairs, we measure the average angle $\phi$ of the two satellites with respect to the host galaxy major axis (see sketch in Figure~\ref{fig:host_axis}). The position angle of the latter is taken from the SDSS exponential and de Vaucouleurs model fits (the fit with the higher likelihood is adopted). The resulting angle of the satellite pairs is shown in Figure~\ref{fig:host_axis}, where we break down the sample into anti-correlated (29) and correlated (12) satellite pairs. The galaxies with anti-correlated velocities show a marked preference for being aligned with their host's major axis, with a mean angle of $\langle \phi \rangle = 35.0\deg \pm 4.4\deg$ (the uncertainties are estimated by bootstrap resampling). In contrast, the satellite pairs with correlated velocities present the opposite behavior, having $\langle \phi \rangle = 57.2\deg \pm 6.1\deg$.

\begin{figure}
\begin{center}
\includegraphics[viewport= 30 30 530 535, clip,width=\hsize]{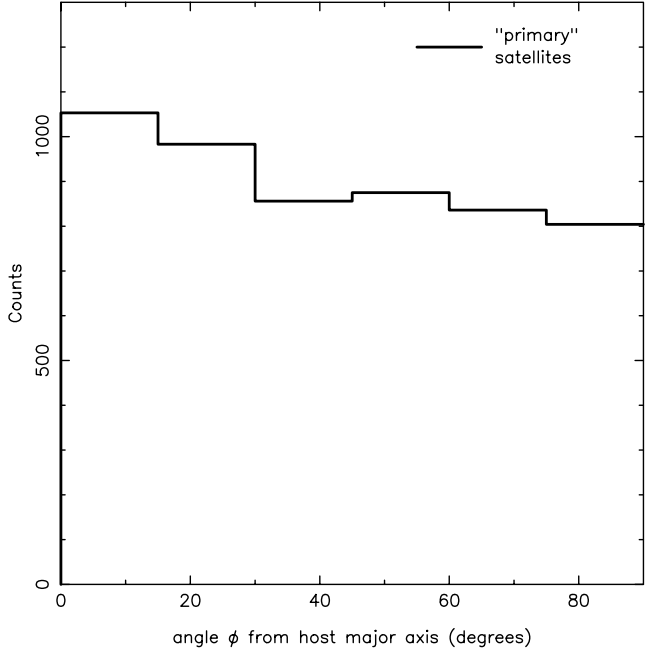}
\end{center}
\caption{$\phi$ distribution of ``primary'' satellites. The ``primary'' satellites defined in this contribution (which are derived from spectroscopic data) also show a preference to be located near the host major axis, possessing a mean angle of $\langle \phi \rangle = 42.50\deg \pm 0.37\deg$. Comparison to Fig.~\ref{fig:host_axis} shows that the effect is much weaker than for the anti-correlated satellite galaxy pairs  reported in Paper~I.}
\label{fig:primaries_host_axis}
\end{figure}

The ``primary'' satellites defined here provide a useful comparison to these samples. In Fig.~\ref{fig:primaries_host_axis} we show the corresponding distribution of angles $\phi$, which is found to peak towards the host major axis, although in a less marked manner than the sample with anti-correlated velocities from Paper~I.

\section{Discussion}
\label{sec:Discussion}

\subsection{Positional correlations}

The analysis presented in Section~\ref{sec:Satellite_satellite} shows that there is a clear enhancement of ``secondary'' candidate satellites (selected using photometric redshifts), diametrically opposite high-probability spectroscopically-confirmed ``primary'' satellites. We also show that this effect exceeds the correlation present in cosmological structure formation simulations at the $3\sigma$ level, once we account for the high contamination in photometric redshift samples. This follows what would be expected if a significant fraction of satellites were aligned in planar structures, as claimed in Paper~I, and reinforces the statistics of the discovery presented in that paper, providing independent confirmation of the ubiquity of satellite spatial correlations in the low redshift ($z<0.1$) universe. This population of satellite systems is evident for satellites within about 150~kpc from their host.

\eject

\subsection{Alignment with host major axis}
\label{sec:Discussion_alignment}
It is not surprising that the distribution of our ``primary'' satellites as a function of angle from the host's major axis (Figure~\ref{fig:primaries_host_axis}) is very similar to the satellite distributions found in other recent SDSS studies (see, e.g. \citealt{2010ApJ...709.1321A}, their Figure~4), using different selection criteria. This raises the question: are the satellite-satellite positional correlations we report here simply a consequence of the well-known positional correlation of satellites with their host's major axis \citep{2004MNRAS.348.1236S,2009MNRAS.395.1184S,2005ApJ...628L.101B,2007MNRAS.376L..43A,2010ApJ...709.1321A}? This question can be answered using Figure~\ref{fig:primaries_host_axis}, which is effectively a probability distribution function for finding a (single) satellite at an angle $\phi$. If the only correlation that satellites have is via their host's major axis (i.e. if they are otherwise statistically independent), then the positions of satellite pairs can be selected by picking two $\phi$ values independently (and at random) from this distribution. Thus we can calculate the consequence of this hypothesis of the independence of satellites by measuring the angles between satellite pairs picked in this way. The resulting distribution of $\alpha$ is shown in Figure~\ref{fig:angle} (blue dashed-line histogram), and is clearly very flat. Just like for the real data, we have ignored the range $90\deg < \alpha<180\deg$, so as to avoid having to consider binary satellites, which are obviously not statistically independent. A Kolmogorov-Smirnov test shows that the observed distribution of ``secondary'' satellites (continuous line histogram) is inconsistent with being formed out of statistically-independent satellites (99.8\% confidence).

Thus we conclude that the driving correlation that these satellites possess is not with the host major axis, but rather with each other. Future work, exploring how the angular momentum of the host correlates with that of the satellite systems, will likely help connect how the properties of the inner galaxy affect the satellite structures.

\subsection{Concerns raised by Cautun et al.}
\label{sec:cautun}

In a recent preprint, C14 raised a series of criticisms of the work presented in Paper~I. In the following subsections we will examine each one of these issues in turn.

\subsubsection{Angular distribution}
\label{sec:angular_distribution}
C14 undertook a similar analysis to that presented in Section~\ref{sec:Satellite_satellite}, examining the angular distribution of galaxies with respect to spectroscopically-confirmed satellites\footnote{The right-hand side region from Fig. 2 of C14 (with angles between 90$\deg$ and 180$\deg$) corresponds to the angles covered in our Fig. 3, the angle definition that we used being the complement of the angle used in their analysis.}. The angles between the confirmed satellites and the neighboring galaxies show similar clustering statistics to the Millennium \citep{Springel:2005gv} and Millennium II simulations, and hence C14 argue that there is no need for an additional disk-like satellite population to account for the SDSS observations. C14 also argue that the model they use to de-contaminate the satellite counts is very accurate, and we find no reason to doubt this. 

The analysis presented above in Section~\ref{sec:Satellite_satellite} came to the opposite conclusion, however. We strongly suspect that the difference between our respective results is due to the use of photometric redshifts in the present work, which by removing a very large fraction of foreground and background contaminants (as we discussed in presenting Figures~\ref{fig:redshift_comparison} and \ref{fig:contamination}), allow the properties of the satellites to be measured. This would explain the strong angular correlation (with $\alpha$) found in Figure~\ref{fig:angle} based on the photometric redshift sample, which is confirmed by the (largely contamination free) spectroscopic sample shown in Figure~\ref{fig:angle_spectro}. 

For this discussion, it is instructive to consider the contamination fraction that the C14 method suffers from. To estimate this, we made a similar selection to C14, choosing candidate satellites to $r=21.0$, and disregarding the photometric redshift information. As for the contamination test performed in Section~\ref{subsec:Contamination}, we shifted the sky positions of the real isolated host galaxies of satellite systems and re-measured the number of (apparent) satellites in the new (incorrect) locations. The hosts were moved by 4 virial radii (chosen in the same absolute-magnitude dependent way as C14) along the four cardinal directions North, South, East and West. We find that for hosts in the magnitude range $-23\le {\rm M_r} \le -20$, the satellite counts at distances between $100<R<150\kpc$ are 95\% contaminated. This fraction increases to 98\% for hosts in the range $-21\le {\rm M_r} \le -20$. 

This is likely the origin of the discrepancy between our analyses. The C14 approach is sensitive to structures close to the hosts, where the density of satellites is significant compared to the background, but at large distance the signal is drowned out by a vast population of contaminants. Any signal present at large distance is further diminished by their choice of amalgamating all measures starting at $20\kpc$.

C14 argue in their Sect.~5 that if 50\% of the satellites were located in a disk (in their self-described ``simple" model), $N_O/N_A$ should be of the order of only $1.1$, which is close to the observed value beyond 100~kpc with no decontamination at all. It is interesting to note that, at distances of $\sim 90\kpc$ from the host, this is in fact the value predicted by MS2 with no contaminants for 0\% of satellites located in a planar structure. So C14 estimate that the observed value should not be corrected for any contamination, which clearly does not make any sense. Concerning their ``simple model", it should be clear that modeling the planar structures of satellites as simple ``disk" structures is not necessarily wise, as the structure seen in, e.g., M31 is certainly not a ``disk". If instead one would model the planar structures as, e.g., dumb-bell structures, it should be immediately clear that the observed $N_O/N_A$ could actually tend towards infinity. Contrary to C14, we therefore conclude that the angular distribution of satellites presented here actually provides strong support for the results of Paper~I.

\subsubsection{Parameter selections}
\label{sec:Parameter_selections}
C14 also argue that the significance of the result presented in Paper~I diminishes if different selection parameters are adopted to define the sample of satellite pairs. Our initial aim in undertaking the Paper~I study was to determine if satellite structures similar to what we had discovered around the M31 galaxy \citep{2013Natur.493...62I} could be found statistically in the SDSS. Our current understanding of the M31 satellite structure motivated almost all of the parameter choices, as detailed in Paper~I. Hence those parameter choices were not arbitrary, but have some chance of reflecting the nature of these structures, based at least on what we infer from M31 (and the Milky Way whose satellite system has similar properties).

The parameter choices they examined were the maximum redshift of the sample, the radial extent, the velocity threshold and the satellite-host magnitude threshold.

\begin{itemize}
\item {\it Maximum redshift.} We initially selected $z=0.05$ as the maximum redshift, as this choice is commonly adopted in other SDSS studies of the nearby universe. As the distance modulus at $z=0.05$ is $m-M=36.75$, one may appreciate that the faintest spectroscopically-observed SDSS galaxies at that redshift (magnitude limit of $r=17.8$) must be intrinsically very bright for a satellite, with absolute magnitudes of $M_r = -19$.\\
In each of the satellite pairs considered in Paper~I, there is of course a brighter and a fainter member. It turns out that of the reported satellite pairs, the pair that has the brightest of the fainter members has $M_r=-19.39$. Hence beyond a redshift of $z=0.06$ we would not find {\it any} of the pairs reported in Paper~I. It is therefore perfectly possible that the falloff in significance with SDSS data beyond the redshift cut of $z=0.05$ reflects the intrinsic properties of the satellite galaxies that partake in these hypothesized alignments.
\item {\it Radial extent.} The choice of considering only a region out to $150\kpc$ in Paper~I was motivated by the size of our survey of Andromeda \citep{2009Natur.461...66M} within which the M31 satellite structure was discovered. Obviously, reducing this radial limit will reduce the sample size, and if the contamination is not reduced much faster, the result will be a lower significance (as is the case). Increasing the radius must come at the cost of greater contamination. However, as we have seen with {\it independent data} in Figure~\ref{fig:ratio}, there is little sign of an angular correlation signal beyond $150\kpc$. So again, it is not surprising that lower significance is found by adopting a maximum radial extent $>150\kpc$. Indeed, this may be actually revealing the typical extent of these structures. 
\item {\it Velocity threshold.} The adopted velocity threshold of $300\kms$ was chosen as this value corresponds to twice the central velocity dispersion of the stellar halo of M31 ($152\kms$, \citealt{Chapman:2006ia}). A two-sigma selection criterion is appropriate for explorations where the contamination is unknown but expected to be moderate.
Since there are very few satellite pairs that did not have anti-correlated velocities in the $\alpha=8\deg$ sample presented in Paper~I, it is not at all surprising that by reducing this limiting velocity value the significance of the detection drops. On the other hand, increasing the velocity limit incurs the chance of including more contaminants, and we suspect that is occurring here (see C14, their Figure 6, though note that their selection procedure is similar, but not identical to that presented in Paper~I). For the case of the M31 satellite alignment, the maximum radial velocity difference of a satellite with respect to M31 is $256.5\kms$, so if all systems were like M31, increasing the velocity threshold beyond $300\kms$ could only decrease the detection significance. 

In their Section 5, C14 argue that our radial extent and velocity thresholds are too conservative, and that by relaxing them, to $250\kpc$ and $500\kms$ respectively, we would incur only minor contamination in our satellite samples. The implication is that our conclusions are incorrect, since the signal decreases when such thresholds are imposed. We checked this claim directly using SDSS data, as follows. We selected isolated host galaxies precisely as in Section 2 above. The number of candidate satellite galaxies between $150\kpc < R < 250\kpc$ and with $300\kms<|v-v_{\rm host}|<500\kms$ around the sample of hosts is $774$. By shifting the hosts to offset locations by 4 virial radii (in a similar way to that described in Section~\ref{sec:angular_distribution} above) and remeasuring the number of surrounding galaxies, we find that the contamination is 100\% (the background count exceeds the true count, presumably because many hosts are on the edges of larger structures). 
To present an alternative test, we also selected hosts that possess a ``primary satellite'', which (as described in Section 2) are located within a projected radius of $150\kpc$. Requiring the presence of a high-confidence ``primary'' satellite means that we are reproducing similar selection biasses as in our alignment analysis above.
The number of candidate satellite galaxies between $150\kpc < R < 250\kpc$ and with $300\kms<|v-v_{\rm host}|<500\kms$ around this new sample of hosts is $390$. Shifting the hosts to offset locations by 2(4) virial radii, we find the corresponding contamination to be 51\%(32\%). The differences in these estimations of the contamination are substantial, and reflect the complex selection biasses involved in dealing with real samples. However, all these estimates indicate that contamination in the spectroscopic sample becomes important if one accepts satellites in a wide range in radius and velocity around their hosts, contrary to the claim by C14.
\item {\it Satellite-host magnitude threshold.} This $\Delta M_r$ parameter is the minimal magnitude difference between the satellite and host, and hence sets the maximum brightness, with respect to the host, that a galaxy may have to be considered a satellite. We feel that considering $\Delta M_r < 1$ makes little sense since then the system is more akin to a binary rather than a host with satellites. However, as one increases $\Delta M_r$, the number of systems that pass the cut drops, and so again it is not at all surprising that the significance of the result drops.
\end{itemize}

\subsubsection{Same-side satellite pairs}

Finally, C14 show a measurement of the angular correlation of pairs of satellites on the same side of the host (rather than on opposite sides). They argue that the lack of velocity correlation rules out the interpretation of the opposite-side pairs presented in Paper~I. 

First, this is obviously not equivalent to showing that the original result on opposite pairs was wrong. There were two main reasons why we did not present such same-side pairs in Paper~I: 
\begin{itemize}
\item Some of the objects will be binaries orbiting each other, so one cannot straightforwardly use the adopted velocity correlation/anti-correlation criterion to ascertain whether they are part of a coherent structure, since their orbital velocity around each other may be large compared to the line of sight component of the orbital velocity of the binary around the host. This is particularly problematic given that the satellite galaxies we observe in the SDSS out to $z=0.05$ are necessarily intrinsically quite massive, and possess a substantial fraction of the mass of their hosts. 
\item By selecting objects that are near each other, one has a higher probability of picking objects in denser environments, such as dwarf galaxies that are part of an infalling group. In the SDSS survey, due to fiber positioning constraints, targets that are closer than $55$" cannot be observed with a single plate \citep{2005AJ....129.2562B}, and this leads to significant incompleteness when galaxies are separated by less than this amount. Hence the kinematics of denser groupings of satellites may be both intrinsically complex and complex to interpret due to incomplete knowledge of the velocities of neighboring galaxies. 
\end{itemize}

Visually checking the same-side detections shows that some of the measurements of {\it velocity anti-correlation} are simply measurements of the center and a bright knot in the spiral arm of a satellite. This means that one needs to impose a minimum separation criterion between satellites to obtain a sensible sample; we estimate $25\kpc$ is a reasonable limit for this purpose. To ensure that the velocity of any binary galaxies around each other is small compared to the orbital velocity around the host, it is also necessary to limit $\Delta M_r$, the magnitude difference between the host and the brightest satellite of the pair. We repeat the analysis of Paper~I taking $\Delta M_r = 2$, which yields a total of 16 same-side pairs: of these, 10 pairs have correlated velocities, 5 pairs have anti-correlated velocities, and we reject one pair after visual inspection, as it is a satellite of a neighboring bright galaxy (present in DR10 of the SDSS but not in the VAGC-DR7).

Thus we find twice as many same-side satellite pairs with correlated velocities than with anti-correlated velocities, which is indeed the sense that would be expected if large galaxies host co-rotating satellite structures as suggested by the analysis presented in Paper~I. However, given the additional uncertainties of using same-side satellites, we refrain from drawing any strong conclusions from this small additional sample of satellite pairs.

\section{Conclusions}
\label{sec:Conclusions}

Spurred by discoveries of satellite alignments in the Local Group and other hosts, we attempted to explore further afield in Paper~I, searching the $z<0.05$ universe with the SDSS survey to examine the incidence of satellite pairs that lie opposite each other across their host and have anti-correlated radial velocities. Using sample selection parameters derived from our M31 observations \citep{2013Natur.493...62I}, we found a strong  ($4\sigma$) enhancement of anti-correlated satellite pairs for satellites that are close to being diametrically opposite each other (within a tolerance angle of $\alpha=8\deg$). This velocity anti-correlation is consistent with the presence of rotating alignments of satellite galaxies, similar to what is present around the Milky Way and M31, and in contradiction with dark-matter-only cosmological simulations. Furthermore, in the immediate ($2\mpc$) environment around the hosts containing the satellite pairs with anti-correlated velocities, we found that neighboring galaxies are strongly correlated ($7\sigma$) with the direction defined by the satellite pairs.

The present work has extended this analysis to consider candidate satellite galaxies that have photometric redshifts consistent with belonging to systems that contain a luminous host and at least one satellite galaxy with a spectroscopically measured redshift. In the raw counts we find a slight preference (17\%) for the photometrically-selected satellite galaxy candidates (in the $100<R<150\kpc$ distance interval) to lie directly opposite a spectroscopically-confirmed satellite, compared to at $90\deg$. However, this corresponds to a factor of $\sim 3$ enhancement after accounting for the numerous contaminants in the photometric redshift sample. Repeating this analysis with spectroscopically-selected satellites only (which are essentially free of contaminants, but probe much more luminous galaxies), we find a similar trend.

To follow the radial profile of the excess population of photometrically-selected satellite candidates at low $\alpha$ values, we calculate the $N_O/N_A$ ratio defined in Figure~\ref{fig:ratio}, and find that this excess extends to $\sim 150\kpc$ (though we note that the sample of ``primary'' satellites that define the angle $\alpha$ only extends to $150\kpc$). The significance of the galaxy overdensity in the ``opposite'' quadrant amounts to $3.3\sigma$. An identical experiment on the Millennium~II cosmological simulation shows a $N_O/N_A$ ratio of virtually unity out to large radius, if the contamination is properly accounted for. We note that tidal dwarf galaxy models \citep{2010A&A...523A..32K} have been argued to produce rotational satellite systems within about $200\kpc$ of their host galaxy \citep{2011A&A...532A.118P}, although a lot of work still has to be done to demonstrate that the lifetime and various characteristics of dwarf spheroidal galaxies, such as multiple stellar populations of different ages and velocity dispersions in different satellites of the same system, mass-metallicity relation, etc., can quantitatively be reproduced in such a model. Preliminary work in this direction includes \citet{2007A&A...470L...5R} and \citet{2014MNRAS.437.3980P} showing that the tidal dwarfs do not destroy themselves, \citet{2015MNRAS.447.2512P} showing that they survive for at least 3$\,$Gyr as self-regulated star-forming systems, and \citet{2015CaJPh..93..169K} showing that the mass--metallicity relation of tidal dwarfs models can match the observed relation.

We also examined the orientation with respect to the host major axis of the satellite pairs reported in Paper~I. Although the statistics are still meagre, the distributions shown in Fig.~\ref{fig:host_axis} suggest that the systems with diametrically opposite satellite pairs with anti-correlated velocities are themselves aligned with the major axis of the host ($\langle \phi \rangle = 35.0\deg \pm 4.4\deg$).

We examined carefully the interesting criticisms of Paper~I raised by \citet{Cautun:2014un} in a recent preprint: they did not uncover an error in our previous analysis, nor demonstrate that it was flawed in any way, but merely argue that the significance of the anti-correlation weakens (or strengthens by taking a smaller maximum redshift) when selection criteria are changed. We discuss how the parameter selection choices adopted in Paper~I were not arbitrary, but rather reflect our knowledge of the systems detected in the Local Group, and that altering these in the manner suggested by C14 has the expected effect on sample size and hence on the significance of the result. We also argue how the satellite sample based on photometric redshifts actually strongly supports our earlier work. Figure~\ref{fig:ratio} shows that the positional angular correlation signal is present before $150\kpc$ but then fades away, which may reveal the actual physical extent of these structures, and explain why the kinematic anti-correlation signal also fades away beyond this radius. We finally discuss why analyses based on same-side satellites are hard to interpret, although with strong quality cuts they are not inconsistent with our earlier results (10/15 same-side pairs uncontaminated by binaries having correlated velocities within a tolerance angle of $8\deg$). 

In summary, we provide here a measurement of the angular correlation of satellite galaxies, which is in excess ($3\sigma$) of expectations from $\Lambda$CDM simulations, adding to the growing evidence for planar satellite structures from Paper~I and the Local Group discoveries: the combined probability of having satellite systems arranged as in the Local Group in dark matter-only $\Lambda$CDM simulations is $<10^{-5}$, a problem which the present $3\sigma$ discrepancy on larger scales thus makes yet more severe. However, the sample sizes are still small (for the spectroscopic-only selection), the contamination is large (for the photometric redshift samples), so larger clean samples are now needed to allow further refinement of these intriguing results.

\acknowledgments

We would like to thank Marius Cautun, Wenting Wang, Till Sawala, and the anonymous referee for their very helpful comments. G.F.L thanks the Australian research council for support through his Future Fellowship (FT100100268) and Discovery Project (DP110100678). The Millennium-II Simulation databases used in this paper and the web application providing online access to them were constructed as part of the activities of the German Astrophysical Virtual Observatory (GAVO).

%\bibliography{ms}
%\bibliographystyle{apj}
%\end{document}

\end{document}